\documentstyle[12pt]{article}
\begin{document}
\pagestyle{empty}
\hspace*{12.4cm}IU-MSTP/36 \\
\hspace*{13cm}hep-lat/9906016 \\
\hspace*{13cm}June, 1999
\begin{center}
 {\Large\bf Axial Anomaly in Lattice Abelian Gauge Theory \\
in Arbitrary Dimensions}
\end{center}

\vspace*{1cm}
\def\thefootnote{\fnsymbol{footnote}}
\begin{center}{\sc Takanori Fujiwara,} 
{\sc Hiroshi Suzuki}
and {\sc Ke Wu}\footnote{On leave of absence from 
Institute of Theoretical Physics, Academia Sinica, P.O.Box 2735, Beijing 
100080, China}
\end{center}
\vspace*{0.2cm}
\begin{center}
{\it Department of Mathematical Sciences, Ibaraki University,
Mito 310-8512, Japan}
\end{center}
\vfill
\begin{center}
{\large\sc Abstract}
\end{center}
\noindent
Axial anomaly of lattice abelian gauge theory in hyper-cubic regular 
lattice in arbitrary even dimensions is investigated by applying the 
method of exterior differential calculus. The topological invariance, 
gauge invariance and locality of the axial anomaly determine the explicit 
form of the topological part. The anomaly is obtained up to a multiplicative 
constant for finite lattice spacing and can be interpreted as the Chern 
character of the abelian lattice gauge theory.

%

\newpage
\pagestyle{plain}


\noindent
Recent discovery of lattice Dirac operator \cite{has,HLN,neub} satisfying 
Ginsperg-Wilson (GW) relation \cite{GW} and implementing exact chiral symmetry 
\cite{lus} have opened up new possibility of understanding nonperturbative 
behaviors of chiral gauge theories on the lattice. The axial anomaly arises as 
the nontrivial Jacobian factor of the fermion measure \cite{fuj} under the chiral 
transformations and is related to the index of the Dirac operator \cite{HLN,lus}. 
Perturbative evaluation of the axial anomaly in the continuum limit 
was carried out in ref. \cite{KY} by using the overlap Dirac operator \cite{neub} 
and the axial anomaly of the contimuum theories were reproduced. See also 
refs. \cite{adam,fuj2}. Such explicit 
analysis becomes rather involved by its own right. However, it is plausable that 
the axial anomaly on the lattice is also related to some topological object as in 
continuum theory and its structure can be determined by invoking the method of 
differential geometry on the discrete lattice. In fact it was argued in ref. 
\cite{lus2} that the topological invariance of the index of the GW Dirac operator 
and the gauge invariance almost fix the form of the axial anomaly, and the 
topologcal part of the anomaly is obtained up to a multiplicative constant 
for finite lattice spacing in four dimensions. 

In this note we investigate the axial anomaly of lattice abelian gauge theory
in euclidean hyper-cubic regular lattice in arbitrary even dimensions  
by applying the method of exterior differential calculus.  We find that the 
topological invariance, gauge invariance and locality of the axial anomaly  
also determine the explicit form of the topological part in arbitrary dimensions.
The axial anomaly is a natural extension of the result obtained in ref. \cite{lus2} 
and has characteristic structure of products of field strengths contracted with 
the Levi-Civita symbol but each argument of the field strengths is shifted so 
that the axial anomaly acquires topological nature. We argue that such shifts 
in the arguments can be naturally understood within the framework of noncommutative 
differential calculus \cite{sit} and the axial anomaly is indeed the Chern character 
of abelian gauge theory on the discrete lattice.


Let us begin with some basic definitions in noncommutative differential 
calculus \cite{sit} on the hyper-cubic regular lattice ${\bf Z}^D$ of unit 
lattice spacing. See also ref. \cite{lus2}. We denote the 
generators of exterior differential algebra by ${\rm d}x_\mu$ ($\mu=1,\cdots,D$). 
They satisfy 
\begin{eqnarray}
  \label{eq:gene}
  {\rm d}x_\mu {\rm d}x_\nu=-{\rm d}x_\nu {\rm d}x_\mu~, \qquad
  f(x){\rm d}x_\mu={\rm d}x_\mu f(x-\hat\mu)~, 
\end{eqnarray}
where $f(x)$ is an arbitrary function on ${\bf Z}^D$. In noncommutative 
differential calculus ${\rm d}x_\mu$ does not commute with the coordinates $x_\mu$ 
as in ordinary differential calculus. Instead, it generates a shift of the 
coordinates in the direction indicated by $\hat\mu$. 

Differential $k$-forms on ${\bf Z}^D$ can be defined by
\begin{eqnarray}
  \label{eq:n-form}
  f=\frac{1}{k!}f_{\mu_1\cdots\mu_k}(x){\rm d}x_{\mu_1}\cdots {\rm d}x_{\mu_k}~,
\end{eqnarray}
where $f_{\mu_1\cdots\mu_k}(x)$ is completely antisymmetric in $\mu_1,
\cdots,\mu_k$. 
The vector space of $k$-forms on ${\bf Z}^D$ is denoted by $\Omega_k$.

On the discrete lattice one can introduce two kind of 
difference schemes, the forward and backward difference operators 
$\partial_\mu$ and $\partial^\ast_\mu$ defined by
\begin{eqnarray}
  \label{eq:fbdiff}
  \partial_\mu f(x)=f(x+\hat\mu)-f(x)~, \qquad
  \partial^\ast_\mu f(x)=f(x)-f(x-\hat\mu).
\end{eqnarray}
Exterior differential operator ${\rm d}:\Omega_k\rightarrow\Omega_{k+1}$ 
on forms can be defined by the forward difference operator as
\begin{eqnarray}
  \label{eq:extderiv}
  {\rm d}f=\frac{1}{k!}\partial_\mu f_{\mu_1\cdots\mu_k}(x)
  {\rm d}x_\mu {\rm d}x_{\mu_1}\cdots {\rm d}x_{\mu_k}~.
\end{eqnarray}
Since two successive differences commute $\partial_\mu\partial_\nu
=\partial_\nu\partial_\mu$, the exetrior differential operator satisfies 
nilpotency relation ${\rm d}^2=0$. This enables us to define closed forms and 
exact forms as in ordinary exterior calculus.

The remarkable property of noncommutative differential calculus is that 
the Leibniz rule for the ordinary exterior differential calculus 
holds true because of the second property in (\ref{eq:gene}). Let $f$ and $g$ be 
$k$- and $l$-forms, then one easily finds
\begin{eqnarray}
  \label{eq:leib}
  {\rm d}(f(x)g(x))={\rm d}f(x) g(x)+(-1)^kf(x){\rm d}g(x) ~.
\end{eqnarray}

We shall use another kind of exterior differential operator, the divergence 
operator ${\rm d}^\ast:\Omega_k\rightarrow \Omega_{k-1}$, defined by
\begin{eqnarray}
  \label{eq:divop}
  {\rm d}^\ast f=\frac{1}{(k-1)!}\partial^\ast_\mu f_{\mu\mu_2\cdots\mu_k}(x)
  {\rm d}x_{\mu_2}\cdots {\rm d}x_{\mu_k}~.
\end{eqnarray}
It also satisfies nilpotency ${\rm d}^{\ast2}=0$. 

The nilpotency of the exterior difference operator naturally leads to 
an analog of Poincar\'e lemma. We quote it here from ref. \cite{lus2}: 

\vskip .3cm
\noindent{\bf Lemma 1} {\sl Let $f$ be a closed $k$-form on ${\bf Z}^D$ with compact 
support and $\sum_x f(x)=0$ for $k=D$, then there exists a $(k-1)$-form $g$ 
such that $f={\rm d}g$.}

\vskip .3cm
\noindent
Since ${\rm d}^\ast$ is also nilpotent, one can state Poincar\'e lemma in the 
following form:

\vskip .3cm
\noindent{\bf Lemma 2} {\sl Let $f$ be a $k$-form on ${\bf Z}^D$ with compact 
support satisfying ${\rm d}^\ast f=0$ and $\sum_x f(x)=0$ for $k=0$, then there 
exists a $(k+1)$-form $g$ such that $f={\rm d}^\ast g$.}
\vskip .3cm
Let us introduce another copy of the lattice ${\bf Z}^D$ and the generators of 
exterior algebra ${\rm d}y_\mu$ ($\mu=1,\cdots,D$) satisfying 
\begin{eqnarray}
  \label{eq:dys}
  && {\rm d}x_\mu {\rm d}y_\nu=-{\rm d}y_\nu {\rm d}x_\mu~, \qquad
  {\rm d}y_\mu {\rm d}y_\nu=-{\rm d}y_\nu {\rm d}y_\mu~, \nonumber \\
  && f(x,y){\rm d}x_\mu={\rm d}x_\mu f(x-\hat\mu,y)~, \qquad
  f(x,y){\rm d}y_\mu={\rm d}y_\mu f(x,y-\hat\mu)~, 
\end{eqnarray}
where $f$ is an arbitrary function of $(x,y)\in{\bf Z}^D\times{\bf Z}^D$. 
We define difference operators by
\begin{eqnarray}
  \label{eq:diffopxy}
  \partial_\mu f(x,y)=f(x+\hat\mu,y)-f(x,y)~,\qquad 
  f(x,y)\overleftarrow\partial_\mu=f(x,y+\hat\mu)-f(x,y)~, \nonumber\\
  \partial^\ast_\mu f(x,y)=f(x,y)-f(x-\hat\mu,y)~,\qquad 
  f(x,y)\overleftarrow\partial{}^\ast_\mu=f(x,y)-f(x,y-\hat\mu)~.
\end{eqnarray}
A differential $(k,l)$-form on ${\bf Z}^D\times{\bf Z}^D$ is defined by 
\begin{eqnarray}
  \label{eq:dklform}
  f=\frac{1}{k!l!}f_{\mu_1\cdots\mu_k|\nu_1\cdots\nu_l}(x,y){\rm d}x_{\mu_1}\cdots
  {\rm d}x_{\mu_k}{\rm d}y_{\nu_1}\cdots {\rm d}y_{\nu_l}~,
\end{eqnarray}
where $f_{\mu_1\cdots\mu_k|\nu_1\cdots\nu_l}(x,y)$ is completely antisymmetric 
in $\mu_1,\cdots,\mu_k$ and in $\nu_1,\cdots,\nu_l$, separately, and is 
assumed to have compact support on ${\bf Z}^D\times\{y\}$ and $\{x\}\times{\bf Z}^D$. 
We denote the vector space of $(k,l)$-forms by $\Omega_{k,l}$. 

The exterior differential with respect to $x$ or $y$ is denoted by ${\rm d}_x$ or 
${\rm d}_y$. For the $(k,l)$-form (\ref{eq:dklform}) they satisfy 
\begin{eqnarray}
  \label{eq:extdxdy}
  && {\rm d}_xf=\frac{1}{k!l!}\partial_\mu f_{\mu_1\cdots\mu_k|\nu_1\cdots\nu_l}(x,y)
  {\rm d}x_{\mu}{\rm d}x_{\mu_1}\cdots
  {\rm d}x_{\mu_k}{\rm d}y_{\nu_1}\cdots {\rm d}y_{\nu_l}~, \nonumber\\
  && {\rm d}_yf=\frac{(-1)^k}{k!l!}f_{\mu_1\cdots\mu_k|\nu_1\cdots\nu_l}(x,y)
  \overleftarrow\partial_\nu 
  {\rm d}x_{\mu_1}\cdots
  {\rm d}x_{\mu_k}{\rm d}y_\nu{\rm d}y_{\nu_1}\cdots {\rm d}y_{\nu_l}~.
\end{eqnarray}
Divergence operators (\ref{eq:divop}) can also be extended to $(k,l)$-forms:
\begin{eqnarray}
  \label{eq:divopxy}
  &&{\rm d}_x^\ast f=\frac{1}{(k-1)!l!}\partial^\ast_\mu 
  f_{\mu\mu_2\cdots\mu_k|\nu_1\cdots\nu_l}(x,y)
  {\rm d}x_{\mu_2}\cdots
  {\rm d}x_{\mu_k}{\rm d}y_{\nu_1}\cdots {\rm d}y_{\nu_l}~, \nonumber\\
  && {\rm d}_y^\ast f=\frac{(-1)^k}{k!(l-1)!} 
  f_{\mu_1\cdots\mu_k|\nu\nu_2\cdots\nu_l}(x,y)\overleftarrow\partial{}^\ast_\nu 
  {\rm d}x_{\mu_1}\cdots
  {\rm d}x_{\mu_k}{\rm d}y_{\nu_2}\cdots {\rm d}y_{\nu_l}~. 
\end{eqnarray}
It is straightforward to show that these operators satisfy nilpotency relations
\begin{eqnarray}
  \label{eq:dxdy}
  {\rm d}_x^2={\rm d}_y^2=({\rm d}_x+{\rm d}_y)^2=0~, \qquad
  {\rm d}_x^{\ast2}={\rm d}_y^{\ast2}=({\rm d}^\ast_x+{\rm d}^\ast_y)^2=0~.
\end{eqnarray}

A differential form $\omega$ satisfying 
\begin{eqnarray}
  \label{eq:dxdyw}
  {\rm d}_x^\ast{\rm d}_y^\ast\omega(x,y)=0
\end{eqnarray}
is of special interest because it gives rise to a sequence of forms related by 
descent equations. Later we encounter $(2m,2)$-forms of this type in analyzing 
axial anomaly in abelian lattice gauge theory. To be definite let us consider a 
$(k,l)$-form $\omega^{k,l}$ satisfying (\ref{eq:dxdyw}), then by the Poincar\'e 
lemma there exist forms $\omega^{k\pm1,l\mp1}$ satisfying 
\begin{eqnarray}
  \label{eq:dxwdyw}
  {\rm d}_x^\ast\omega^{k+1,l-1}(x,y)+{\rm d}_y^\ast\omega^{k,l}(x,y)=0~, \qquad
  {\rm d}_x^\ast\omega^{k,l}(x,y)+{\rm d}_y^\ast\omega^{k-1,l+1}(x,y)=0~.
\end{eqnarray}
Since these forms also satisfy (\ref{eq:dxdyw}), they lead to new forms 
$\omega^{k\pm2,l\mp2}$. Such procedure can be continued until one ends up with 
$\omega^{k+l,0}$ and $\omega^{0,k+l}$ for $k+l\leq D$ or $\omega^{D,k+l-D}$ and 
$\omega^{k+l-D,D}$ for $k+l> D$ . If we define a formal sum of differential 
forms 
\begin{eqnarray}
  \label{eq:sumofforms}
  \omega=\sum_{j=0}^{\min\{k+l,D\}}\omega^{k+l-j,j}~, 
\end{eqnarray}
then the descent equations can be compactly expressed as
\begin{eqnarray}
  \label{eq:unidec}
  ({\rm d}_x^\ast+{\rm d}_y^\ast)\omega(x,y)=0~.
\end{eqnarray}
We restrict ourselves to $m\equiv k+l\leq D$ and solve the descent equations. 
As we shall see, only the forms with $m\leq D$ appear in analyzing the axial anomaly.

We first define an $m$-form $\alpha^{m}$ on ${\bf Z}^D$ by 
\begin{eqnarray}
  \label{eq:a0m}
  \alpha^{m}(y)&\equiv&\sum_{x}\omega^{0,m}(x,y) \nonumber\\
  &\equiv&\frac{1}{m!}\alpha_{\nu_1\cdots\nu_m}(y){\rm d}y_{\nu_1}\cdots{\rm d}y_{\nu_m}
\end{eqnarray}
This can be shown to be divergence free 
\begin{eqnarray}
  \label{eq:diva0m}
  {\rm d}^\ast_y\alpha^m(y)=0
\end{eqnarray}
by the descent equation ${\rm d}^\ast_y\omega^{0,m}(x,y)
=-{\rm d}^\ast_x\omega^{1,m-1}(x,y)$. 
The key observation in solving the descent equations
is that $\omega^{0,m}(x,y)$ can be decomposed as
\begin{eqnarray}
  \label{eq:decompw0m}
  \omega^{0,m}(x,y)=\alpha^{0,m}(x,y)+{\rm d}^\ast_x\vartheta^{1,m}(x,y)~,
\end{eqnarray}
where we have introduced $\alpha^{0,m}(x,y)\equiv\delta_{x,y}\alpha^m(y)$ with 
$\delta_{x,y}$ the Kronecker $\delta$-symbol and 
$\vartheta^{1,m}(x,y)$ is some form in $\Omega_{1,m}$. This can be shown by 
applying the Poincar\'e lemma for $\omega^{0,m}(x,y)-\alpha^{0,m}(x,y)$ as a
0-form on ${\bf Z}^D\times\{y\}$. The 
descent equation ${\rm d}^\ast_x\omega^{1,m-1}+{\rm d}^\ast_y\omega^{0,m}=0$ 
can then be solved as follows
\begin{eqnarray}
  \label{eq:solvdesceq}
  {\rm d}^\ast_x\omega^{1,m-1}&=&-{\rm d}^\ast_y\omega^{0,m} \nonumber \\
  &=&-{\rm d}^\ast_y\alpha^{0,m}(x,y)+{\rm d}^\ast_x{\rm d}^\ast_y
  \vartheta^{1,m}(x,y) \nonumber \\
  &=&{\rm d}^\ast_x\alpha^{1,m-1}(x,y)+{\rm d}^\ast_x{\rm d}^\ast_y
  \vartheta^{1,m}(x,y)~,
\end{eqnarray}
where we have defined a $(1,m-1)$-form $\alpha^{1,m-1}(x,y)$ by
\begin{eqnarray}
  \label{eq:a1m-1}
  \alpha^{1,m-1}(x,y)&\equiv&\frac{1}{(m-1)!}\alpha_{\mu\nu_1\cdots\nu_{m-1}}(x)
  \delta_{x,y-\hat\mu}{\rm d}x_\mu{\rm d}y_{\nu_1}\cdots{\rm d}y_{\nu_{m-1}}~.
\end{eqnarray}
The key relation ${\rm d}^\ast_y\alpha^{0,m}(x,y)
=-{\rm d}^\ast_x\alpha^{1,m-1}(x,y)$ can be seen as follows. First note that 
it can be explicitly written as
\begin{eqnarray}
  \label{eq:conc}
  &&\frac{1}{(m-1)!}(\delta_{x,y}\alpha_{\mu\nu_1\cdots\nu_{m-1}}(y)
  -\delta_{x,y-\hat\mu}\alpha_{\mu\nu_1\cdots\nu_{m-1}}(y-\hat\mu))
  {\rm d}y_{\nu_1}\cdots{\rm d}y_{\nu_{m-1}} \nonumber\\
  &&=-\frac{1}{(m-1)!}(\alpha_{\mu\nu_1\cdots\nu_{m-1}}(x)
  \delta_{x,y-\hat\mu}-\alpha_{\mu\nu_1\cdots\nu_{m-1}}(x-\hat\mu)
  \delta_{x-\hat\mu,y-\hat\mu})
  {\rm d}y_{\nu_1}\cdots{\rm d}y_{\nu_{m-1}}~.
\end{eqnarray}
This holds true if one notices the divergence free condition (\ref{eq:diva0m}), i.e.,
\begin{eqnarray}
  \label{eq:df}
  \partial_\mu^\ast\alpha_{\mu\nu_1\cdots\nu_{m-1}}(x)
  =\alpha_{\mu\nu_1\cdots\nu_{m-1}}(x)-\alpha_{\mu\nu_1\cdots\nu_{m-1}}(x-\hat\mu)
  =0~.
\end{eqnarray}

We see from (\ref{eq:solvdesceq}) that $\omega^{1,m-1}$ can be expressed as
\begin{eqnarray}
  \label{eq:w1m-1}
  \omega^{1,m-1}(x,y)=\alpha^{1,m-1}(x,y)
  +{\rm d}^\ast_x\vartheta^{2,m-1}(x,y)+{\rm d}^\ast_y
  \vartheta^{1,m}(x,y)
\end{eqnarray}
for some form $\vartheta^{2,m-1}$ in $\Omega_{2,m-1}$. This procedure of solving 
the descent equations can be carried out until all the froms 
$\omega^{j,m-j}$ $(j=0,\cdots,m)$ are obtained. One can easily convince himself that 
$\omega^{k,l}$ is given by
\begin{eqnarray}
  \label{eq:wkl}
  \omega^{k,l}(x,y)=\alpha^{k,l}(x,y)+{\rm d}^\ast_x\vartheta^{k+1,l}(x,y)
  +{\rm d}^\ast_y\vartheta^{k,l+1}(x,y)~,
\end{eqnarray}
where $\vartheta^{k+1,l}(x,y)
\in\Omega_{k+1,l}$ and $\vartheta^{k,l+1}(x,y)\in\Omega_{k,l+1}$ are some 
forms on ${\bf Z}^D\times{\bf Z}^D$ and $\alpha^{k,l}(x,y)$ is defined by
\begin{eqnarray}
  \label{eq:akl}
  \alpha^{k,l}(x,y)&\equiv&\frac{1}{k!l!}\alpha_{\mu_1\cdots\mu_k\nu_1\cdots\nu_l}(x)
  \delta_{x,y-\hat\mu_1-\cdots-\hat\mu_k}{\rm d}x_{\mu_1}\cdots{\rm d}x_{\mu_k}
  {\rm d}y_{\nu_1}\cdots{\rm d}y_{\nu_l}~.
\end{eqnarray}
It is straightforward to show that $\alpha^{k,l}$ satisfy the descent equations 
(\ref{eq:dxwdyw}).


We now turn to the analysis of axial anomaly of abelian lattice gauge 
theory on ${\bf Z}^D$. We shall use gauge potential $A(x)=A_\mu(x){\rm d}x_\mu$ 
instead of link variable $U_\mu(x)=\exp iA_\mu(x)$. The field strength 
$F(x)={\rm d}A(x)=\frac{1}{2}F_{\mu\nu}(x){\rm d}x_\mu{\rm d}x_\nu$ can be 
identified with the usual definition 
$-i\ln [U_\mu(x)U_\nu(x+\hat\mu)U_\mu(x+\hat\nu)^{-1}U_\nu(x)^{-1}]$  
if field configurations satisfy 
\begin{eqnarray}
  \label{eq:admiss}
  \sup_{x,\mu,\nu}|F_{\mu\nu}(x)|<\epsilon \qquad
  \Biggl(0<\epsilon<\frac{\pi}{3}\Biggr)~.
\end{eqnarray}
We assume that the gauge field configurations are subject to this 
admissibility condition \cite{lus2}. 

The axial anomaly $q(x)$ is expressed by the GW Dirac operator
$D(x,y)\equiv D(x)\delta_{x,y}$ as \cite{HLN,lus}
\begin{eqnarray}
  \label{eq:lxanom}
  q(x)&=&{\rm tr}\:\gamma_5\Biggl[1-\frac{1}{2}D(x)\Biggr]\delta_{x,x}~.
\end{eqnarray}
It is assumed to depend locally on the gauge potential and is 
gauge invariant. Furthermore, the sum of $q(x)$ over the lattice is the index of 
the GW Dirac operator and is topologically invariant. This implies  under an 
arbitrary local variation of gauge potential $\delta A_\mu(x)$ 
\begin{eqnarray}
  \label{eq:dq}
  \sum_x \delta q(x)&=&0 ~.
\end{eqnarray}
These properties largely restrict the structure of the axial anomaly as argued in 
ref. \cite{lus2}. In fact the generic structure of the axial anomaly is obtained 
in four dimensions. This result can be extend to arbitrary even dimensions $D=2n$. 
We state this in the following theorem:

\vskip .5cm\noindent{\bf Theorem} {\sl Let $q(x)$ be a gauge invariant function which 
depends locally on the gauge potential and the sum over the lattice is 
a topological invariant, then it can be expressed as
\begin{eqnarray}
  \label{eq:theorem}
  q(x)&=&\Biggl(\frac{1}{2}\Biggr)^n
  C\epsilon_{\mu_1\nu_1\cdots\mu_n\nu_n}F_{\mu_1\nu_1}(x)
  F_{\mu_2\nu_2}(x+\hat\mu_1+\hat\nu_1) \nonumber\\
  &&\times\cdots\times
  F_{\mu_n\nu_n}(x+\hat\mu_1+\hat\nu_1+\cdots+\hat\mu_{n-1}+\hat\nu_{n-1})
  +\partial_\mu^\ast k_\mu(x)~,
\end{eqnarray}
where $C$ is a constant, $\epsilon_{\mu_1\cdots\mu_D}$ is 
the Levi-Civita symbol in $D$ dimensions and $k_\mu(x)$ is 
a gauge invariant current. }

\vskip .5cm\noindent
Before turning to the proof of this theorem, we give here a few remarks. We first note
the connection of the topological term $q(x)-\partial^\ast_\mu k_\mu(x)$ with the 
noncommutative differential calculus. The shifts of the arguments of the field 
strengths in (\ref{eq:theorem}) are very important. It can naturally be 
interpreted as the Chern character of abelian gauge theory on the lattice
\begin{eqnarray}
  \label{eq:chern}
  C(F(x))^n&=&\Biggl(\frac{1}{2}\Biggr)^n
  C\epsilon_{\mu_1\nu_1\cdots\mu_n\nu_n}F_{\mu_1\nu_1}(x)
  F_{\mu_2\nu_2}(x+\hat\mu_1+\hat\nu_1) \nonumber \\
  &&\times\cdots\times
  F_{\mu_n\nu_n}(x+\hat\mu_1+\hat\nu_1+\cdots+\hat\mu_{n-1}
  +\hat\nu_{n-1}){\rm d}^{D}x~,
\end{eqnarray}
where ${\rm d}^{D}x\equiv {\rm d}x_1\cdots{\rm d}x_{D}$ is the 
volume form and the arguments of $F_{\mu\nu}$ are shiftted by 
the noncommutativity nature (\ref{eq:gene}). Secondly, the anomaly coefficient $C$ 
can be found easily from the well-known results 
in continuum theories \cite{fuj2}. In the case of abelian 
chiral gauge theory the gauge 
anomaly also takes the form (\ref{eq:theorem}) and should be cancelled 
to guarantee the 
gauge invariance. In anomaly free theories the topological term is 
absent but the second 
term in (\ref{eq:theorem}) may be nonvanishing. It represents an 
artificial breaking of 
the gauge symmetry due to the finite lattice spacing and can be 
removed by a suitable choice of the path integral measure \cite{lus3}. 

To verify the theorem let us consider a gauge invariant $2m$-form $\alpha^{2m}$ on 
${\bf Z}^{D}$ which depends locally on the gauge potential 
and satisfy ${\rm d}^\ast\alpha^{2m}=0$, then its coefficients can be expressed as
\begin{eqnarray}
  \label{eq:alpmexp}
  \alpha_{\mu_1\nu_1\cdots\mu_m\nu_m}(x)=\beta_{\mu_1\nu_1\cdots\mu_m\nu_m}+
  \sum_y\omega_{\mu_1\nu_1\cdots\mu_m\nu_m|\rho}(x,y)A_\rho(y)~,
\end{eqnarray}
where $\omega_{\mu_1\nu_1\cdots\mu_m\nu_m|\rho}(x,y)$ 
are the coefficients of a gauge invariant $(2m,1)$-form $\omega^{2m,1}(x,y)$
defined by (\ref{eq:dklform}) and are explicitly given by 
\begin{eqnarray}
  \label{eq:expformomega}
  \omega_{\mu_1\nu_1\cdots\mu_m\nu_m|\rho}(x,y)&=&\int_0^1dt\Biggl(
  \frac{\partial \alpha_{\mu_1\nu_1\cdots\mu_m\nu_m}(x)}{\partial A_\rho(y)}
  \Biggr)_{A\rightarrow
    tA}~.
\end{eqnarray}
The $\beta_{\mu_1\nu_1\cdots\mu_m\nu_m}$ is the field independent part of 
$\alpha_{\mu_1\nu_1\cdots\mu_m\nu_m}(x)$. The 
dependences of $\beta_{\mu_1\nu_1\cdots\mu_m\nu_m}$ on the lattice coordinates 
can be excluded by assuming the translation invariance of $q(x)$, i.e., it 
depends on the lattice coordinates through the gauge potential or the coordinate 
differences. 

The gauge invariance of $\alpha^{2m}$ and the divergence free condition 
${\rm d}^\ast\alpha^{2m}=0$ imply 
\begin{eqnarray}
  \label{eq:gianddiv}
  {\rm d}^\ast_x\omega^{2m,1}={\rm d}^\ast_y\omega^{2m,1}=0~.
\end{eqnarray}
By the Poincar\'e lemma we see that there exist a form $\omega^{2m,2}(x,y)$ 
such that
\begin{eqnarray}
  \label{eq:omegm2}
  \omega^{2m,1}(x,y)&=&-{\rm d}^\ast_y\omega^{2m,2}(x,y)~, \qquad
  {\rm d}^\ast_x{\rm d}^\ast_y\omega^{2m,2}(x,y)\:\:=\:\:0~.
\end{eqnarray}
The first of these relations implies that (\ref{eq:alpmexp}) can be 
expressed by the field strength as 
\begin{eqnarray}
  \label{eq:alpmexp2}
  \alpha_{\mu_1\nu_1\cdots\mu_m\nu_m}(x)=\beta_{\mu_1\nu_1\cdots\mu_m\nu_m}+
  \frac{1}{2}\sum_y\omega_{\mu_1\nu_1\cdots\mu_m\nu_m|\rho\sigma}(x,y)F_{\rho\sigma}(y)~,
\end{eqnarray}
while the second one leads to the decomposition of $\omega^{2m,2}(x,y)$ 
as in (\ref{eq:wkl})
\begin{eqnarray}
  \label{eq:dcompwm2}
  \omega^{2m,2}(x,y)=\alpha^{2m,2}(x,y)+{\rm d}^\ast_x\vartheta^{2m+1,2}(x,y)
  +{\rm d}^\ast_y\vartheta^{2m,3}(x,y)~.
\end{eqnarray}
The forms $\alpha^{2m,2}$, $\vartheta^{2m+1,2}$ and 
$\vartheta^{2m,3}$ are all gauge invariant \cite{lus2} and in particular $\alpha^{2m,2}$ 
can be found from (\ref{eq:akl}) for $k=2m$ and $l=2$ with some gauge invariant 
$(2m+2)$-form $\alpha^{2m+2}$. Note that $\alpha^{2m+2}$ is also divergence free as 
is $\alpha^{2m}$. 
It is now straightfoward 
to show that $\alpha_{\mu_1\nu_1\cdots\mu_m\nu_m}(x)$ is given by
\begin{eqnarray}
  \label{eq:am2}
  \alpha_{\mu_1\nu_1\cdots\mu_m\nu_m}(x)&=&\beta_{\mu_1\nu_1\cdots\mu_m\nu_m}+
  \frac{1}{2}\alpha_{\mu_1\nu_1\cdots\mu_m\nu_m\rho\sigma}(x)F_{\rho\sigma}
  (x+\hat\mu_1+\hat\nu_1+\cdots+\hat\mu_m+\hat\nu_m) \nonumber\\
  &&+\partial_\mu^\ast\Biggl(\frac{1}{2}
  \sum_y\vartheta_{\mu\mu_1\nu_1\cdots\mu_m\nu_m|\rho\sigma}(x,y)F_{\rho\sigma}(y)
  \Biggr)~.
\end{eqnarray}
The field independent constant 
term can be cast into a total divergence for $m<n$ by introducing
\begin{eqnarray}
  \label{eq:tam1}
  \bar\beta_{\mu\mu_1\nu_1\cdots\mu_m\nu_m}(x)&=&x_{[\mu}
  \beta_{\mu_1\nu_1\cdots\mu_m\nu_m]}~,
\end{eqnarray}
where $[\cdots]$ stands for the antisymmetrization of indices. The case $m=n$ is 
special. In this case 
$\beta_{\mu_1\nu_1\cdots\mu_n\nu_n}$ is proportional to 
$\epsilon_{\mu_1\nu_1\cdots\mu_n\nu_n}$. Furthermore, 
we can not convert it to a total divergence of some $(D+1)$-form. 
We thus obtain
\begin{eqnarray}
  \label{eq:rec}
  \alpha_{\mu_1\nu_1\cdots\mu_m\nu_m}(x)&=&\frac{1}{2}
  \alpha_{\mu_1\nu_1\cdots\mu_m\nu_m\rho\sigma}(x)F_{\rho\sigma}
  (x+\hat\mu_1+\hat\nu_1+\cdots+\hat\mu_m+\hat\nu_m) \nonumber \\
  &&+\partial_\mu^\ast\bar\vartheta_{\mu\mu_1\nu_1\cdots\mu_m\nu_m}(x) \qquad\hbox{for}
  \qquad m<n~, \nonumber\\
  \alpha_{\mu_1\nu_1\cdots\mu_n\nu_n}(x)&=&C\epsilon_{\mu_1\nu_1\cdots\mu_n\nu_n}
  \qquad \hbox{for} \qquad m=n~, 
\end{eqnarray}
where $\bar\vartheta_{\mu\mu_1\nu_1\cdots\mu_m\nu_m}(x)$ is given by
\begin{eqnarray}
  \label{eq:vartheta}
  \bar\vartheta_{\mu\mu_1\nu_1\cdots\mu_m\nu_m}(x)&\equiv&
  \frac{2m+1}{D-2m}\bar\beta_{\mu\mu_1\nu_1\cdots\mu_m\nu_m}(x)
  +\frac{1}{2}
  \sum_y\vartheta_{\mu\mu_1\nu_1\cdots\mu_m\nu_m|\rho\sigma}(x,y)F_{\rho\sigma}(y)~.
\end{eqnarray}
It is completely antisymmetric in its indices and is gauge invariant.
Eq. (\ref{eq:rec}) serves as a recurrence relation in 
expanding $q(x)$ in terms of the gauge potential. 

We now turn to the proof of the theorem. The general argument given 
above allow us to expand $q(x)$ as 
\begin{eqnarray}
  \label{eq:expq0}
  q(x)&=&\frac{1}{2}\alpha_{\mu\nu}(x)F_{\mu\nu}(x)
  +\partial_\lambda^\ast\bar\vartheta_\lambda(x)~,
\end{eqnarray}
where a 2-form $\alpha^2\equiv\frac{1}{2}\alpha_{\mu\nu}{\rm d}x_\mu{\rm d}x_\nu$  
satisfies ${\rm d}^\ast\alpha^2=0$ as can be easily seen from the 
topological invariance (\ref{eq:dq}). So we can apply (\ref{eq:rec}) also 
for $\alpha_{\mu\nu}(x)$. This leads to 
\begin{eqnarray}
  \label{eq:expq1}
  q(x)&=&\Biggl(\frac{1}{2}\Biggr)^2
  \alpha_{\mu_1\nu_1\mu_2\nu_2}(x)F_{\mu_1\nu_1}(x)
  F_{\mu_2\nu_2}(x+\hat\mu_1+\hat\nu_1) \nonumber\\
  && +\partial^\ast_\lambda\Biggl(\bar\vartheta_\lambda(x)+
  \frac{1}{2}\bar\vartheta_{\lambda\mu\nu}(x)F_{\mu\nu}(x+\hat\lambda)\Biggr)~.
\end{eqnarray}
In deriving this we have used the relation
\begin{eqnarray}
  \label{eq:theF}
  \partial^\ast_\lambda(\bar\vartheta_{\lambda\mu\nu}(x)F_{\mu\nu}(x+\hat\lambda))
  =\partial^\ast_\lambda\bar\vartheta_{\lambda\mu\nu}(x)F_{\mu\nu}(x)
  +\bar\vartheta_{\lambda\mu\nu}(x)\partial_\lambda F_{\mu\nu}(x)
\end{eqnarray}
and the Bianchi identity $\partial_{[\lambda}F_{\mu\nu]}=0$. 
Again a 4-form $\alpha^4\equiv\frac{1}{4!}\alpha_{\mu\nu\rho\sigma}
{\rm d}x_\mu{\rm d}x_\nu{\rm d}x_\rho{\rm d}x_\sigma$ 
satisfies ${\rm d}^\ast \alpha^4=0$, and the current appearing in the 
second line of (\ref{eq:expq1}) is gauge invariant.
Obviousely, this procedure can be repeated until we end up with 
(\ref{eq:theorem}). This completes the proof of the theorem. 

In conclusion we have obtained the axial anomaly of lattice abelian 
gauge theory by extending the arguments of ref. \cite{lus2} to arbitrary 
even dimensions. It is given by the Chern character of abelian gauge 
theory on the lattice. This result can also be achieved by the systematic 
BRST analysis \cite{fsw1} extended to lattice gauge theory. In our present treatment 
the role of noncommutativity nature (\ref{eq:gene}) is somewhat indirect. But in the 
BRST analysis noncommutative differential calculus plays an essential role.

\end{document}